# High level architecture evolved modular federation object model[*]

WANG Wenguang[1], XU Yongping[1], CHEN Xin[2,3], LI Qun[1], WANG Weiping[1]

1. College of Information System and Management, National University of Defense Technology, Changsha 410073, P. R. China
2. Beijing Key Laboratory of Intelligent Telecommunications Software and Multimedia,
   Beijing University of Posts and Telecommunications, Beijing 100876, P. R. China
3. State Key Lab for Complex Systems Simulation Beijing 100101, P. R. China

**Abstract:** To improve the agility, dynamics, composability, reusability, and development efficiency restricted by monolithic Federation Object Model (FOM), a modular FOM was proposed by High Level Architecture (HLA) Evolved product development group. This paper reviews the state-of-the-art of HLA Evolved modular FOM. In particular, related concepts, the overall impact on HLA standards, extension principles, and merging processes are discussed. Also permitted and restricted combinations, and merging rules are provided, and the influence on HLA interface specification is given. The comparison between modular FOM and Base Object Model (BOM) is performed to illustrate the importance of their combination. The applications of modular FOM are summarized. Finally, the significance to facilitate composable simulation both in academia and practice is presented and future directions are pointed out.

## 1. Introduction

High Level Architecture (HLA)[1-5] is a common simulation framework to support the interoperability and reusability of various simulation applications. It has been widely used for acquisition, training, and testing in defense community.

Driven by the extension of application scope, the development of new technology, and the need of net-centric simulation in Global Information Grid (GIG)[6], many deficiencies of HLA on interoperability, extensibility, and reusability have been revealed during the past decade. One prominent deficiency is that although Federation Object Models (FOMs) contain all the information to be exchanged in federation (i.e., multiple simulations), the monolithic and static architecture defined by FOM restricts the agility, dynamics, composability, reusability, and the development efficiency (e.g., FOM cannot be expanded and composed dynamically after creating a federation).

Additionally, to facilitate the reusability of simulation resources and rapid development of applications, composable simulation[7,8] has drawn significant attention in modeling and simulation community. HLA, being the leading standard in distributed simulation community, also needs to be extended so as to improve the composability itself. Although Base Object Model (BOM)[9] facilitates the composability of HLA framework at the conceptual level, FOM should also be improved at the implementation level. Meanwhile, as provided in the IEEE standard[1-5], HLA needs to be reviewed and revised every five years. Hence, Simulation Interoperability Standards Organization (SISO), being an IEEE standard development organization, established HLA Evolved Product Development Group (PDG) to revise HLA[10,11]. As one of the major enhancements, modular FOM[12-18] divides FOM into small pieces that can be composed and expanded to form the common information exchange model dynamically. It can improve the agility, dynamics, composability, reusability, and the development efficiency. Furthermore, it can also facilitate dealing with real time and uncertain decision or application integration

[*] This project was supported by the National Natural Science Foundation of China (60674069, 60574056)

---

problems in a highly dynamic and agile sphere, such as net-centric wargame. Moreover, it provides a new idea of dynamic and rapid composition and simulation on demand.

For the above reasons, it is necessary to summarize the progress in HLA Evolved modular FOMs to facilitate the research, development, and application of new generation HLA standards and composable simulation. This paper reviews recent progress in HLA Evolved modular FOM, compares modular FOM with BOM, and summarizes the applications of modular FOM. Finally, conclusions are taken and future directions are pointed out.

## 2. Background of HLA and FOM

HLA core standards consist of HLA framework and rules[1], Object Model Template (OMT)[2] and federate interface specification[3]. Additionally, there are two standards about systems engineering methodology: Federation Development and Execution Process (FEDEP)[4] and Federation Verification, Validation and Accreditation[5]. OMT defines HLA object models into two forms: FOM and Simulation Object Model (SOM). SOM specifies the types of information that an individual federate could provide to or receive from other federates in HLA federation. FOM is the most important part in HLA, which is the common information exchange contact of all federates to meet valid interoperation. FOM uses a common and standard information exchange model for federates. FOM contains identification, object classes and attributes, interactions and parameters, transportation types, synchronization points, and data types [16].

Apart from FOM and SOM, HLA defines Management Object Model (MOM) to monitor and manage Run Time Infrastructure (RTI), federates, and federation.

## 3. Progress in modular FOM

This section summarizes the progress in modular FOM. New standards in this paper refer to the IEEE1516-200x (i.e., x stands for the approved year by IEEE) serial standards developed by HLA Evolved PDG. The old standards refer to IEEE1516-2000 serial standards.

**3.1 Modular concepts introduced in HLA Evolved**

Three important modular concepts are introduced in HLA Evolved, new concepts are added, and some previous are modified[15,16].

Definition 1: **FOM/SOM Module**[15,16] is a subset of the FOM/SOM that contains some or all OMT tables. A FOM/SOM module shall contain complete or scaffolding definitions for all object classes and interaction classes that are superclasses of object classes and interaction classes in the same FOM/SOM module.

Definition 2: **MOM and Initialization Module (MIM)** [15]: A subset of the FOM that contains OMT tables that describe the HLA MOM. The MOM and Initialization Module shall also contain additional predefined HLA constructs, such as object and interaction roots, data types, transportation types, and dimensions. HLA provides a standard MOM and Initialization Module. The user may also supply a MOM and Initialization Module that contains extensions.

Definition 3: **Scaffolding object/interaction description**[15,16] is a description of an object/interaction class in a FOM/SOM module that has a name identical to the name of a regular object class description provided by another FOM/SOM module and that provides no additional properties such as publish/subscribe indicator or attributes/parameters. Regular object/interaction class description contains at least the object/interaction class name and publish/subscribe indicator.

Based on the above modular concepts, the definitions of FOM and SOM are extended so that FOM/SOM can be composed by corresponding modules. To specify the composite FOM and composite FOM Document Data (FDD) built by modular FOMs, current FOM and current FDD concepts are added in new standards:

Definition 4: **Current FOM**[15,16] is the union of the FOM Modules and one MOM and Initialization Module that have been specified in the creation of the federation execution or by any federate that has joined the federation execution. The sum operation is carried out according to the rules as prescribed in IEEE 1516.2-200x. If all FOM modules have been provided the Current FOM is equal to the FOM and before this had happened, the Current FOM is a true subset of the FOM."

The definition of current FDD is similar to that of current FOM. The concepts of identical FOM module and repeated description are added to new standards. Refer to Ref. [15] for details.

The idea of modular FOM can be illustrated by the following example[12] in Fig.1. HLAObjectRoot comes from MIM, rectangles stand for the entity classes from FOM module 1, the other shape stands for those from FOM module 2. MIM and FOM module 1 and 2 can be composed to build the FOM of a federation. The pseudo codes of FOM modules 1, 2 and the composite FOM are shown in Fig.2. The scaffolding definitions of object/interaction classes can be replaced by regular ones.

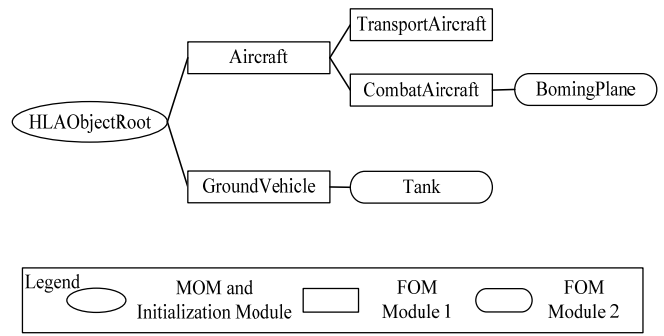

Fig.1 Composite FOM of modular FOMs

| **FOM Module 1**<br>HLAobjectRoot<br>　　Aircraft(Name,Speed,Height)<br>　　　　TransportAircraft(Capability)<br>　　　　CombatAircraft(AmmunitionType)<br>　　Groundvehicle(Name,Speed,Position) | **Composite FOM**<br><br>HLAobjectRoot(PrivilegeToDeleteObject)<br>　　Aircraft(Name,Speed,Height)<br>　　　　TransportAircraft(Capability)<br>　　　　CombatAircraft(AmmunitionType)<br>　　　　　　BombingPlane(BombNumber)<br>　　Groundvehicle(Name,Speed,Position)<br>　　　　Tank(Type) |
|---|---|
| **FOM Module 2**<br>HLAobjectRoot<br>　　Aircraft // scaffolding definition<br>　　　　CombatAircraft // scaffolding definition<br>　　　　　　BombingPlane(BombNumber)<br>　　Groundvehicle // scaffolding definition<br>　　　　Tank(Type) | |

Fig.2 Pseudo codes of composite FOM and modular FOMs

### 3.2 Overall impact on HLA serial standards

The serial standards in IEEE 1516 are interrelated. By that the modular FOM concepts are likely to have impacts on all the core standards. Because FEDEP and federation verification, validation and accreditation are out of the scope of HLA Evolved PDG, modular FOM concepts are not introduced into these two standards.

**(1) Impact on HLA framework and rules**

In the new HLA framework[15], the definitions in subsection 3.1 applies and rule 1 and rule 6 are modified and supplemented. FOM can be specified using one or more FOM modules and one MIM, SOM can be specified using one or more FOM modules and one optional MIM.

**(2) Impact on OMT**

In the new HLA OMT[16] background, the rationale of OMT is modified so that all discussions of HLA object model apply to FOM/SOM, FOM/SOM module, unless explicitly stated otherwise, and FOM/SOM can be specified using one or more modules.

In the new HLA OMT components, it is stated that FOM/SOM module fully conforms to all of the rules and constraints in new OMT specification that, in turn, is a compliant object model.

In object model identification table, References field is extended to support modular FOM/SOM. The semantics of References is not changed. It relates to the pointers to additional information. However, if the object model is intended to represent a FOM/SOM module or a composition of modules, this field shall be extended by Type and Identification subfields. Type subfield specifies the form of the reference material, where Standalone, Dependency and Composed From predefined values may be used to represent independent, dependent, and composed modules. Other values may be used to indicate the general type of reference source. Dependent FOM modules contain references to other FOM modules (such as object classes or data type definitions). Independent or standalone modules can be used without other FOM modules and can only have reference to MIM. Identification subfield specifies Not Available (NA), the names of all the FOM/SOM modules which this module depends on, or the names of all FOM/SOM modules that have been merged to form the current object model respectively if the values in Type subfield are "Standalone", "Dependency" or "Composed From".

The most significant enhancements in the new OMT are FOM/SOM merging rules and principles. They indicate which FOM extensions are allowed and how to merge modules into FOM. The details are discussed in the following two subsections.

**(3) Impact on federate interface specification**

Besides some new definitions, the enhancements and changes focus on how to inform RTI about FOM extensions, how to manage the loaded modules and how to inform federates about FOM extensions. The details are discussed in the later subsection.

### 3.3 Extension principles and merging processes

**(1) Extension principles**

The context of FOM extensions is the following: After federation created according to initial common information exchange model (FOM) or during the execution process, some joined federates indicate how to extend the FOM to exchange some additional information that the other federates do not care about. FOM is a federation-wide common information exchange model. If FOM extensions only apply to local federates, other federates may receive unexpected handles that would require more intrusive changes to HLA interface specification and require a lot of work in RTI to maintain the inconsistencies between the federates. Hence, FOM extensions should be a federation-wide concept[18]. As for the permitted FOM extensions, there are four possible choices[18]:

(a) Allow the addition of new object/interaction classes only at the root level.

(b) Allow (a), and the addition of subclasses to existing classes.

(c) Allow (a) and (b), and the addition of new attributes to existing classes.

(d) Allow (a), (b), and (c), and the addition of new parameters to interaction classes.

Option (a) is the simplest case easy to be implemented, but it cannot extend existing classes because of having limited extensibility. Option (b) is more powerful than (a). Although (b) has not the ability to add new attributes or parameters as (c) and (d) do, these functionalities can be equally implemented by directly adding new subclasses with extended attributes/parameters to existing classes. Furthermore, (b) has a little impact on the instances of existing classes as new extensions are only constrained in new subclasses. Hence, (b) is a better choice in that case. Option (c) increases extensibility, but there is no rule to indicate the ownerships of attributes extensions in existing class instances. So, the maintenance to the extensions adds more complexity to RTI. Option (d) is the most powerful and the hardest to be implemented. Regarding the objects, RTI performs filtering before delivering an attribute update. However, regarding the interactions, it can deliver all associated parameters of interaction classes. Federates that do not know the FOM extensions may receive unexpected parameters. Thus, higher requirements on RTI are needed so as to maintain consistency between the federates. Comparatively, option (b) is the best choice and has been accepted and applied in new HLA standards.

**(2) Merging principles and processes**

The key idea of a merging process of the modular FOMs is that FOM can be regarded as a stateful object. Current FOM reflects the state of FOM. The merging process of modular FOMs is defined as[16]: the first FOM module directly merged into the current FOM. Then, subsequent FOM modules should be performed using a top-down method beginning with the object/interaction classes at the root nodes. All candidate object/interaction classes, attributes/parameters, and associated data types, dimensions, transportations and notes should be considered and aligned. Additional items such as synchronizations, updateRates and notes should be compared and considered for integration into FOM.

The merging principles and processes of metadata, classes, data types, dimensions, synchronization points, transportation type, updateRates, and notes are listed in the annex part of new OMT standard[16]. Classes can be taken as an example.

The process of merging classes consists of two phases: the process of finding duplicate classes and the process of merging into the FOM. The aim of the former is to ignore any duplicates already in FOM, following the principles and process[16]: Compare the class name of the FOM module being inserted with that in current FOM to see if an equivalent already exists; If find duplicate, determine the identical match of the two classes; Compare every sub element of the class respectively from the FOM module and current FOM; Compare datatypes as well.

Unique class names found in a FOM module are considered candidates being merged into current FOM. Steps in merging process include[16]: Check to see if an identical ancestry for the candidate class already exists. Determine compatibility of the candidate and judge if any candidate class parents also need to be merged; Compare the candidate datatype to see whether it already exists in current FOM.

### 3.4 Allowed and disallowed combinations and merging rules

**(1) Allowed and disallowed combinations**

On the basis of extension and merging principles and the concepts of standalone and dependent FOM modules, the allowed and disallowed combinations are illustrated in Fig.3[12]. The principles also apply to SOM modules except that MIM is optional.

**(2) Merging rules**

Merging rules determine the regulations followed by the merging process and the composite FOM[12], for example, an exact equivalency between tables (e.g., switches, time representation, and user supplied tags); Then, duplicates should be equivalent in the union of table elements; duplicates in the union of hierarchical elements should be equivalent or scaffolding; FOM developers are responsible for completing the left metadata in object model identification table except References field. The merge results and rules are provided in Ref. [16].

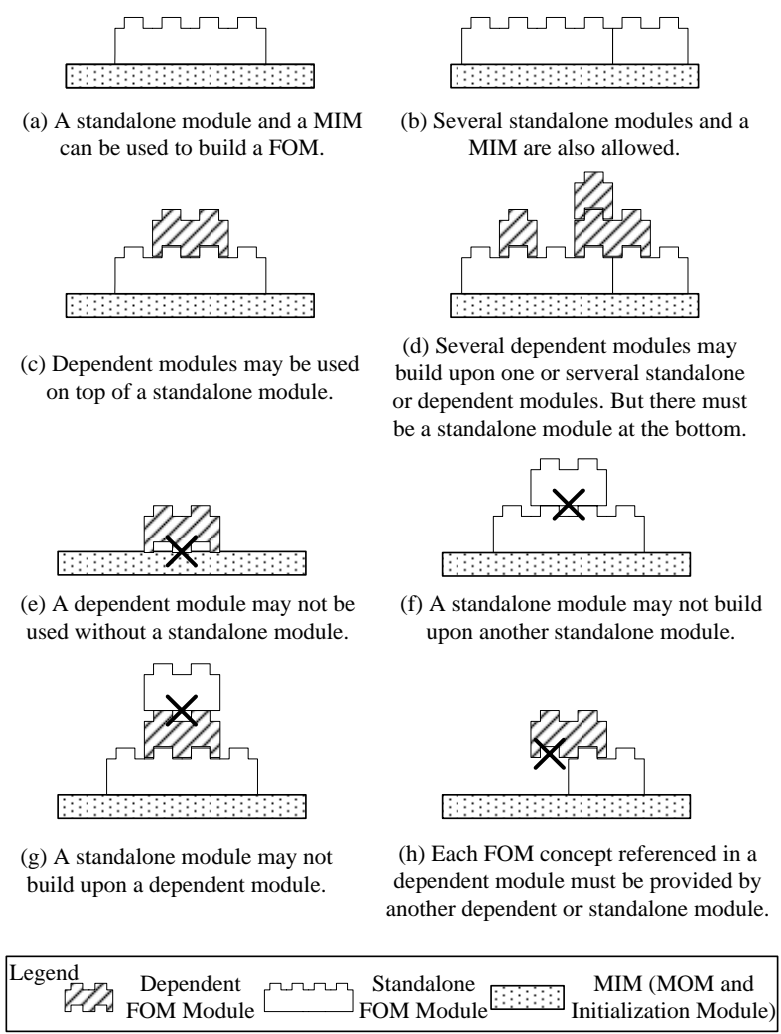

(a) A standalone module and a MIM can be used to build a FOM.

(b) Several standalone modules and a MIM are also allowed.

(c) Dependent modules may be used on top of a standalone module.

(d) Several dependent modules may build upon one or serveral standalone or dependent modules. But there must be a standalone module at the bottom.

(e) A dependent module may not be used without a standalone module.

(f) A standalone module may not build upon another standalone module.

(g) A standalone module may not build upon a dependent module.

(h) Each FOM concept referenced in a dependent module must be provided by another dependent or standalone module.

Legend: Dependent FOM Module / Standalone FOM Module / MIM (MOM and Initialization Module)

**Fig.3 Allowed and disallowed FOM module combinations**

### 3.5 Revisions in federate interface specification

Modular FOM is supported in the new federate interface specification from the following three aspects:

**(1) FOM modules can be specified to RTI when federation or federates joins are created.**

This subsection answers the question how to inform RTI about FOM extensions. The Create Federation Execution and Join Federation Execution services are modified to support list of FOM modules. The FDD provided by federates when they join can include only the FDD extension or some of the current FOM's classes, attributes, and interactions[12,18]. The load operations of FOM modules are atomic. If FOM modules cannot be successfully combined due to incompatible modules or confliction with merging principles or rules, all FOM modules will be rejected throwing an exception and the whole load operation will fail. Once loading is done successfully, MOM will reflect the update of the current FOM. The lifecycle of FOM modules will span the entire life of the federation execution.

**(2) FOM modules are managed by MOM**

This subsection answers the question how to manage the loaded FOM modules. For the HLAfederation class, the HLAFOMmoduleDesignatorList attribute is added to load the entire list of identifiers of all FOM modules into federation execution. Furthermore, the HLAMIMDesignator attribute stores the identifier of MIM. HLAcurrentFDD attribute keeps the current FDD. For HLAfederate class, HLAFOMmoduleDesignatorList attribute is added to load the list of identifiers of all FOM modules by this federate in Join Federation Execution call[12,18].

**(3) The content of FOM modules can be retrieved by MOM interactions.**

This subsection answers the question how to inform federates about FOM extensions. There are two solutions. One of them is adding a new FederateAmbassador callback and the other is letting MOM provide the information. Because FOM extensions are only concerned by late-joined or a few federates (e.g., data logger and management federate), the latter solution is more reasonable. Interactions in MOM are extended in new standard[12,17]. For the federation, the HLArequestFOMmoduleData and HLArequestMIMData interactions are added to present the request to the content of the specified FOM module or MIM respectively. HLAreportFOMmoduleData and HLAreportMIMData interactions present the response to the federation. For federates, HLArequestFOMmoduleData interaction is added to present the request to the content of specified FOM module. HLAreportFOMmoduleData interaction responses to the request from federate.

## 4. Relationships between modular FOM and BOM

Similar to modular FOM, BOM also reflects the idea of breaking conceptual model, simulation object model, or federation object model down into small pieces. There are many differences between them, though. Their relationships and combination deserve further research.

### 4.1 Introduction to BOM

BOM is a component-based simulation object specification developed by SISO. The purpose of BOM is to improve the composability, reusability, and interoperability at the conceptual model level. BOM is defined as "a piece part of a conceptual model, simulation object model, or federation object model, which can be used as a building block in the development and/or extension of a simulation or federation."[9] BOM consists of model identification (metadata), conceptual model definition, model mapping, object model definition, notes, and lexicon. The BOM standards include template specification[9] and guide for BOM use and implementation[19].

### 4.2 Comparison and contrast between BOM and modular FOM

Referring to the literature[9,13,16,19,20], the comparison and contrast between BOM and modular FOM are illustrated in Table 1. The purpose of the two standards is to improve the reusability and composability of HLA models, simulations, or federations that are based on the modular and component idea. BOM focuses on the conceptual aspect, whereas modular FOM on the implementation aspect. The common and unique properties make them complement and combine each other. According to the new HLA OMT standard, BOM can represent one possible method for formulation FOM/SOM module[16]. Chase et al.[21] propose to break the FOM down into more manageable smaller BOMs. The assembling BOMs can be then split into agile FOMs. Bowers et al.[22]

compose and decompose FOM applying BOM idea and envision groups of frequently used BOMs comprising a FOM module.

**Table 1 Comparisons between BOM and Modular FOM**

| Items | BOM | Modular FOM | Similarity and relationship |
|---|---|---|---|
| Standard | SISO standard | IEEE standard | (1) Both based on the assumption that models, simulations or federation can break down into components or modules which can be reused and composed. (2) Both are reusable, dividable and composable. (3) Both based on open standard, both use XML to facilitate extensibility and reusability. (4) Both conform to OMT specification. (5) Share some common structures (HLA object/interaction classes, attributes/parameters, identification table, datatypes notes, lexicon) which facilitate the conversation to each other. (6) Model mapping in BOM can identify reusable FOM modules and facilitate their conversation. (7) BOM can be a modular FOM if it contains information about HLA (e.g. dimensions). But this would limit the BOM's reuse potential for non-HLA implementations. (8) Modular FOM can be a BOM if it has not conceptual model or model mapping. But reusability would be limited. (9) If use BOM to define conceptual model and map to modular FOMs, then it would facilitate the use of both BOM and FOM module, combining conceptual and implementa- |
| Essence | Conceptual model components with mappings to object models | Function pieces of the entire common federation information exchange model | |
| Purpose | Provide component oriented model framework at the conceptual level to facilitate interoperability, reusability and composability of simulation models | Break FOM down into composable modules. Facilitate the agility, dynamics, composability, reusability and development efficiency. | |
| Usage | Describe and share conceptual model to facilitate interoperability and reusability in different simulation frameworks; provide model mappings converting to object model in subsequent process; provide the base for executive software codes | Developed by different communities in different domains; specified to RTI to form FOM; current FOM can be extended by later FOM modules | |
| Concern | Conceptual modeling | Common information exchange model, focuses on implementation | |
| Stakeholder | Federation developer and component designer who want to develop reusable and composable federation component models | Federation developer and end-user who want plug-and-play federation. | |
| Levels of interoperability | Target at the conceptual level | Syntactic level | |
| Interoperability | The conceptual model part in BOM can refer to other BOMs. Model mapping can across conceptual to implementation models of other BOMs, FOMs or other architectures models. Pattern of interplay and state machine can represent the interactions between object models. Higher interoperability. | FOM module can refer to complete or scaffolding definitions of parent classes. FOM module can be used as the basis for other module's further extensions. | |
| Reusability | Conceptual model. Higher reusability. Can be reused across federations or other simulation frameworks besides HLA. BOM can be reused, evolved and applied to specific domains. | Implementation model. Relatively lower reusability. Only reused in HLA. Support implementation of federates conforming to the specification and merging rules. | |
| Extensibility | Easy to extend BOM elements based on XML | Easy to extend elements based on XML | |
| Composability | Several BOMs can form BOM assembly and further be mapped to federates and federation. | FOM/SOM modules can be composed to lager FOM/SOM modules or FOM/SOM | |
| Coupling | Cannot be loaded by RTI. Loose coupling with RTI and implementation (e.g. not have MOM, route and dimensions that tight coupled with RTI); loose coupling between BOMs | Can be loaded and used by RTI. Tight coupling with RTI and implementation; tighter coupling between FOM modules that have extension relationships (such as classes or data) | |
| Structure | Has conceptual model (pattern of interplay, state machine, entity and event type) and model mapping that modular FOM has not. | Has HLA dimensions, time, tags, synchronizations, transportations and switches etc. that BOM has not. | |
| Conceptual model | Describe and share conceptual model in specification manner | Not Available | tion model and supporting the whole process of FEDEP. (10) BOM assembly of several BOMs can be mapped to modular FOMs by mechanisms such as XSLT and using MIM to generate SOM, FOM or aggregate model. (11) Both apply well in HLA framework. |
| Implementation model | Model mapping facilitates the conversation to implementation model. Object model part defines the interface to implement federates or federation | Detailed implementation of object models in particular object/interaction classes, attributes/parameters and runtime information | |
| Model mapping | Entity and event mapping from conceptual model to implementation model | Not Available | |
| Application in FEDEP | Mostly used in the early steps of FEDEP. Reuse general conceptual knowledge for applicable domains. The utility of BOM becomes weaker from step 2 to step 4 in FEDEP (perform conceptual analysis, design federation, and develop federation). | Mostly used in the late steps of FEDEP. Compose functional modules of federates into federation. The utility of modular FOM becomes stronger from step 2 to step 4 in FEDEP. | |

## 5. Use cases of modular FOM

Modular FOM introduces many new opportunities and use cases[14]:

**(1) Use cases in reference FOM**

On one hand, modular FOM can support modular development of reference FOM[23]. FOM modules can be developed by different organizations in different communities and domains. This can improve the efficiency and flexibility, and reduce cost of development. On the other hand, modular FOM can separate concerns and manage extensions of reference FOM effectively.

**(2) Improving the reusability of federation agreement components**

FOM is the common information exchange protocol of the whole federation. Modular FOM can facilitate dividing monolithic federation agreement into small manageable, reusable components that will promote the reusability of modules across federations.

**(3) Supporting long-running federations**

Modular FOM enables new extensions that can be added without repeating the whole process of development and execution. Federates with new capabilities can be added without shutting down the entire federation. It is possible to have GIG style long-running online federations.

At present, many commercial RTI corporations including Pitch and MAK are playing an active role in revising new HLA standard and developing new versions of RTI. MAK has released its own commercial product that supports modular FOM[24,25]. Because HLA Evolved new standards have not been approved by IEEE, MAK RTI 3.2 configures RTI Initialization Data file to support modular FOM[24].

## 6. Conclusion and future work

As the needs of simulation interoperability, reusability, and composability extend continually, simulation systems are moving towards standardization, introduction of components, hierarchy, networks and services abilities[26].

From the viewpoint of academic value, modular FOM divides monolithic federation model into small manageable, reusable, and composite modules. That reflects the idea of modularization. It improves the agility, dynamics, composability, reusability, and the efficiency of development and execution. It also facilitates the research and development of new composable HLA simulation interoperability standards and provides an applicable framework to the simulation community.

From the viewpoint of practical value, modular FOM can facilitate the research and application of new-generation RTI products and associated supporting tools. Dynamic extensions of FOM require after-action review and management federate to improve their functionality. Combining modular FOM with BOM can gain the most benefits of both conceptual model and implementation model. This, in turn, enables a better composition of HLA framework. Modular FOM can change the patterns of federation design and development to a certain degree. It continuously facilitates simulation federations by a dynamic extension and composition of current and future FOM modules. It can also facilitate dealing with real time and uncertain decision, and application integration problems in a highly dynamic and agile sphere such as net-centric wargame. Moreover, it provides a new idea of dynamic and rapid composition and simulation on demand. It promotes the transformation of current simulation resources and the development of new applications, and has important research value and wide application prospect.

Although the research on modular FOM has made great progress, there are still many unsolved problems and difficulties which can be regarded as further research directions:

**(1) Design and implementation of Modular FOM**

The design and implementation of Modular FOM are the most important issues encountered by stakeholders, especially commercial off the shelf RTI corporations. We are doing some exploratory work using open source RTI and referring to MAK RTI.

**(2) Modification of HLA serial standards associated with modular FOM**

Modular FOM impacts not only HLA framework and rules, OMT and interface specification, but also other associated standards, such as FEDEP, federation verification, validation and accreditation, and BOM. These standards should also be revised to support modular FOM.

**(3) Combination with BOM**

Models are the core of simulation. BOM and modular FOM both target at improving the composability of HLA simulation framework. How to combine them effectively, convert each other and automatically generate SOM and FOM need further research.

**(4) Research on how to combine modular FOM with other new technologies in HLA Evolved**

From the development of HLA Evolved, some new improvements[11,27] promote the composability, extensibility, interoperability, reusability, and reliability of HLA such as Web Service API, XML Schema, HLA Evolved fault-tolerant federations, Dynamic Link Compatible APIs, Smart Update Rate. How to combine modular FOM with these new technologies needs further research.

**(5) Improving the composability and interoperability levels of modular FOM**

According to the levels of conceptual interoperability model[28,29], to get meaningful simulation systems, the merging rules of modular FOM should not be constrained at syntactic level, but developed toward semantic, pragmatic, and conceptual levels to reach higher levels of composability.

**(6) Development and application of associated tools and products**

The old HLA standards should be transformed by HLA Evolved modular FOM. Associated RTI products, underlying implementation mechanisms, after-action review, management federate and other previous and novel applications should also evolve for the purpose of modular FOM.

**(7) The combination between modular FOM and other composable simulation approaches and frameworks**

Apart from the common library, product line, interoperability protocol, object model approaches surveyed by Weisel[7], there are many others such as DEVS[30,31], Ontology[32], Model Driven Architecture[33,34] and Service Oriented Architecture approaches[31,34]. In particular, the research on modular FOM within the emerging service-oriented simulation paradigm looks very promising. The relationships, interoperability, and reusability between these frameworks and modular FOM are worth further research.

**WANG Wenguang** was born in 1978. He currently is a Ph.D. candidate. He is a member of SCS, SISO and CASS (Chinese Association for System Simulation). He also is an invited reviewer for the international journal *Simulation Modelling Practice and Theory*. His research interests include service-oriented simulation, simulation composability, and interoperability, etc. E-mail: wgwangnudt@gmail.com

---------------------------------------------------------------------------------------------------------------------

You are welcome to refer to the authors' other publications:

[1] Wang W.G., Wang W.P., Zander J., Zhu Y.F., 2009. Three-dimensional conceptual model for service-oriented simulation. *Journal of Zhejiang University SCIENCE A*, **10**(8):1075-1081. http://www.zju.edu.cn/jzus/2009/A0908/A090801.htm
http://www.springerlink.com/content/p217u140771hx35q/?p=281e44773f0b47d09220f5d992b913c9&pi=2

[2] Wang W.G., Tolk A., Wang W.P., 2009. The levels of conceptual interoperability model: Applying systems engineering principles to M&S. Spring Simulation Multiconference (SpringSim'09). San Diego, CA, USA. http://arxiv.org/abs/0908.0191

[3] Wang W.G., Yu W.G., Li Q., Wang W.P., Liu X.C., 2009. Service-oriented high level architecture. *European Simulation Interoperability Workshop*. 08E-SIW-022. http://arxiv.org/abs/0907.3983

Best kind regards
Wenguang Wang
(wgwangnudt@gmail.com )